\documentclass[12pt]{iopart}
\usepackage{graphicx}
\usepackage{epsfig}
\begin{document}

\title{Observation of a level crossing in a molecular nanomagnet using 
implanted muons}

\author{T. Lancaster$^{1}$, J.S. Moeller$^{1}$, S.J. Blundell$^{1}$,
F.L. Pratt$^{2}$, P.J. Baker$^{2}$, T. Guidi$^{2}$, G.A. Timco$^{3}$ and
R.E.P. Winpenny$^{3}$}
\ead{t.lancaster1@physics.ox.ac.uk} 
\address{$^{1}$Oxford University Department of Physics, Clarendon Laboratory,  Parks
Road, Oxford, OX1 3PU, UK}

\address{$^{2}$ISIS Facility, Rutherford Appleton Laboratory,  Chilton, 
Oxfordshire OX11 0QX, UK}

\address{$^{3}$School of Chemistry and Photon Science Institute, University of Manchester, Oxford Road, Manchester M13 9PL, UK}

\begin{abstract}
We have observed an electronic energy level crossing in a molecular
nanomagnet (MNM) using muon-spin relaxation. This effect, not observed
previously despite several muon studies of MNM systems, provides
further evidence that the spin relaxation of the implanted muon is
sensistive to the dynamics of the electronic spin.  Our measurements
on a broken ring MNM
[H$_{2}$N$^{\mathrm{t}}$Bu$^{\mathrm{is}}$Pr][Cr$_{8}$CdF$_{9}$(O$_{2}$CC(CH$_{3}$)$_{3}$)$_{18}$]
(hereafter Cr$_{8}$Cd), which contains eight Cr ions, show clear
evidence for the $S=0 \rightarrow S=1$ transition that takes place at
$B_{\mathrm{c}}=2.3$~T. The crossing is observed as a resonance-like
dip in the average positron asymmetry and also in the muon-spin
relaxation rate, which shows a sharp increase in magnitude at the
transition and a peak centred within the $S=1$ regime.
\end{abstract}

\maketitle

Molecular nanomagnets (MNMs) \cite{gatteschi} comprise clusters of
transition metal ions. Strong exchange coupling between these ions
within a single molecule results in each molecule possessing a ground
state described by a total spin eigenvalue $S$.  Excited states will
possess other values of $S$ and the splitting of these levels in an
applied magnetic field often leads to level crossings in which an
excited spin state in zero field becomes the new ground state at fields
above the field at which a crossing occurs.  MNMs have been widely
studied in recent years, most recently in anticipation of their
possible deployment as elements of quantum computers \cite{ardavan},
although much interest also centres on the quantum tunnelling of the
magnetization (QTM) which can take place when the magnetic energy
levels are at resonance \cite{gatteschi}. When implanted muons were
first 
used to probe the spin dynamics in these systems,
it was
hoped that they would be sensitive to level crossings.  However, early
studies failed to observe any signature of such crossings \cite{tom1}
and, despite the possible observation of effects ascribed to
a matching of the MNM electronic energy level splitting with that of the muon
hyperfine levels \cite{corti}
and the observation of crossings in broadly related
systems \cite{graf},
 the observation of a crossing in a MNM has
remained elusive until now. Here we demonstrate that muons are
sensitive to the electronic energy level crossings in MNMs. Our
measurements on a broken ring system, made using the new HiFi
spectrometer at the ISIS facility, demonstrate the effect of the level
crossing on the integrated positron asymmetry and on the muon-spin
relaxation rate.

\begin{figure}
\begin{center}
\epsfig{file=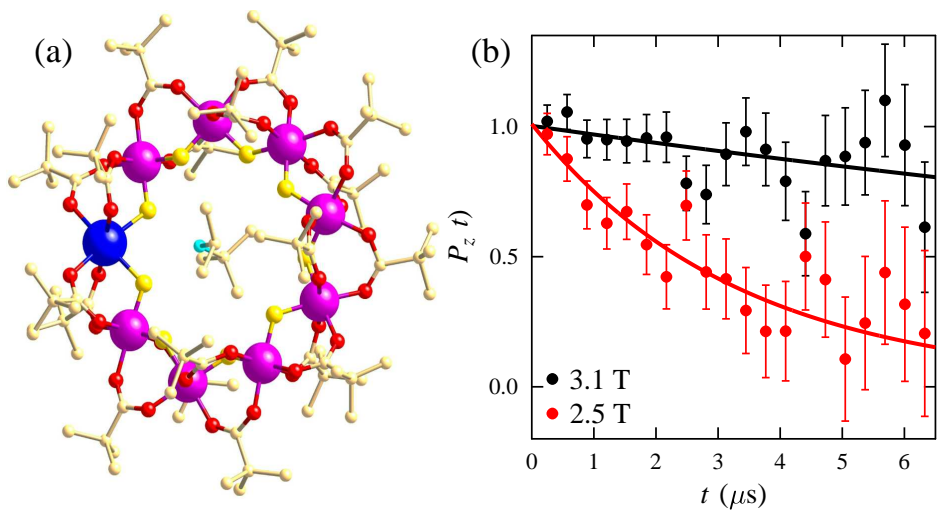,width=14cm}
\caption{(a) The Cr$_{8}$Cd molecule.
(b) Example muon-spin relaxation spectra measured at 70~mK in applied longitudinal magnetic field.
\label{data}}
\end{center}
\end{figure}

The material measured in this study is related to the octonuclear
system [Cr$_{8}$F$_{8}$(O$_{2}$CC(CH$_{3}$)$_{3}$)$_{16}$]
\cite{timco}.  That material has a $S=0$ ground state due to
antiferromagnetic coupling ($J_{\mathrm{Cr-Cr}}\approx 16.9$~K)
between the eight nearest neighbour Cr$^{3+}$ ($s=3/2$) spins.  In
contrast, the broken ring system Cr$_{8}$Cd \cite{timco2} (full
formula [H$_{2}$N$^{\mathrm{t}}$Bu$^{\mathrm{is}}$Pr]
[Cr$_{8}$CdF$_{9}$(O$_{2}$CC(CH$_{3}$)$_{3}$)$_{18}$], shown in
figure~\ref{data}(a)) has one $s=0$ Cd$^{2+}$ ion added to the ring
which interrupts the strong intraring exchange interactions,
effectively disconnecting two Cr$^{3+}$ spins and changing the
topology of the magnetic interactions \cite{furukawa}. This has the
effect of significantly altering the bulk magnetic behaviour of the
system, whose first level crossing ($S=0 \rightarrow S=1$) occurs at a
magnetic field of $B_{\mathrm{c}}=2.3$~T (compared to 7.3~T for
Cr$_{8}$).

In a muon-spin relaxation ($\mu^{+}$SR) experiment \cite{steve},
spin-polarized positive muons are stopped in a target sample. The time
evolution of the muon spin polarization is probed via the positron
decay asymmetry function $A(t)$ to which it is proportional. Our
$\mu^{+}$SR measurements were made on the new HIFI spectrometer
\cite{salman,james} at the ISIS facility, Rutherford Appleton Laboratory,
UK. This instrument allows the application of magnetic fields of up to
$B=5$~T, longitudinal to the initial muon spin direction and is
optimised for time-differential muon spin relaxation studies at a
pulsed muon source.  For the measurements, six crystallites of
Cr$_{8}$Cd, prepared as reported previously \cite{timco2}, were
arranged on a silver plate attached to the cold-finger of a dilution
refrigerator.  The crystallites were aligned such that their $a$-axes
were directed perpendicular to the direction of the applied field.

Example data measured at $T=70$~mK are shown in figure~\ref{data}(b) for two
values of applied magnetic field. The muon polarization $P_{z}(t)$
(which is
proportional to the positron asymmetry) is
seen to decrease monotonically and is well described by an exponential
relaxation function. This is typical behaviour for these MNM systems
\cite{tom},
and can be attributed to dynamic fluctuations of the local magnetic
field distribution
at the muon sites in the material \cite{hayano}. 
In order to 
follow the behaviour probed by the muon across the level crossing in
Cr$_{8}$Cd we plot the
time-averaged asymmetry in figure~\ref{data}(b) for scans in
applied field at temperatures of 70~mK and 20~K. Resonance-like
minima are clearly observable, which may be identified with the
electronic energy level
crossing between $S=0$ and $S=1$ ground states. At 70~mK the minimum
occurs at $B=2.29$~T with a FWHM of approximately $0.4$~T. Increasing the temperature causes the
resonance to broaden in an asymmetric fashion and shifts the
minimum to a slightly higher field of 2.35~T.

\begin{figure}
\begin{center}
\epsfig{file=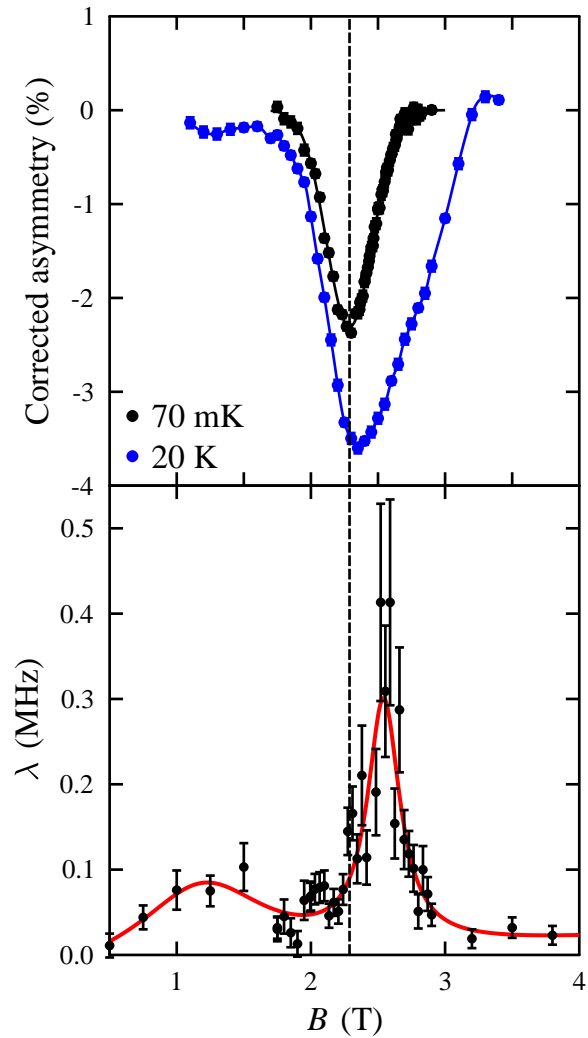,width=9cm}
\caption{(a) Average asymmetry (corrected for background)  as a function of applied magnetic field
  measured at 70~mK and 20~K. Resonance-like minima are observed at the level
crossing. 
(b) Relaxation rate $\lambda$ at 70~mK resulting from fitting the measured
spectra to equation~(\ref{fitting}). The peak is displaced to slightly
higher fields than the level crossing. 
\label{results}}
\end{center}
\end{figure}

Another method of examining the resonance is to fit the
time-differential spectra to the functional form
\begin{equation}
A(t) = A_{\mathrm{rel}} e^{-\lambda t} +A_{\mathrm{bg}}, 
\label{fitting}
\end{equation}
where $A_{\mathrm{rel}}$ is the relaxing amplitude and
$A_{\mathrm{bg}}$ is the background contribution, which
we expect to be highly field-dependent due to the effect of the magnetic
field on the incoming muons and outgoing positrons due to the Lorentz
force. The amplitude $A_{\mathrm{rel}}$ was held fixed throughout the fitting
procedure and the extracted relaxation rate $\lambda$, measured at 70~mK is
shown in figure~\ref{results}(b). The relaxation rate is seen to
increase sharply around $B_{\mathrm{c}}$ and peak at $B=2.54$~T. Note that the
peak is observed well within the $S=1$ regime. (The origin of the
apparent broad,
low-amplitude peak in the low-field region is unclear, although this most-likely represents a
background contribution to the relaxation.)

Our previous study of MNM systems\cite{tom} identified the mechanism
through which the muon spin is relaxed in these materials.
Specifically, measurements on Cr$_{8}$ and on the related $S=1$ MNM
system Cr$_{7}$Mn showed that the muon spin ensemble is relaxed by static
nuclear magnetism in $S=0$ systems such as Cr$_{8}$ and by the large electronic spin
in $S \neq 0$ MNMs such as Cr$_{7}$Mn.  Moreover, a large difference
in relaxation rates between protonated and deuterated samples
demonstrates that the proton fluctuations are largely responsible for
the dephasing of the large MNM electronic spin that we detect with
muons at low temperatures.  It is likely, therefore, that for our
level crossing measurement of Cr$_{8}$Cd, the channels through which
the muon spins are relaxed change quite dramatically from weak nuclear
relaxation in the $S=0$ regime to strong electronic relaxation upon
traversing the level crossing to the $S=1$ regime above
$B_{\mathrm{c}}$. Although it is probable that the fluctuation rate of
the net moment of a molecule is symmetrically peaked about the
crossing, the effective coupling of the muon to the electronic
spins on the molecule is likely to be smoothly turned on upon crossing into
the $S=1$ regime and this may cause the peak
in the muon response to be shifted to slightly higher fields, as we
observe. Another possibility for the shift is that the electronic
fluctuation
rate lies outside the muon time window close to the transition, but
slows above the crossing causing the maximum in $\lambda$ as it 
decends into the regime in which the muon is sensitive.

In conclusion, muon-spin relaxation has been shown to be sensitive to
the level crossing in the molecular nanomagnet Cr$_{8}$Cd. 
This opens up possibilities for its use in probing such crossings in
other systems. Future work will involve
examining the crossings between two $S \neq 0$ states in order to 
further examine the nature of the coupling of the muon to the
molecules. 

Part of this work was carried out at the ISIS facility, Rutherford Appleton
Laboratory, UK and is supported by a beamtime award from STFC (UK). 
This work is supported by the EPSRC (UK) and the EU (MolSpinQIP).

\end{document}